\documentstyle[12pt,a4,epsfig]{article}
\newcommand{\be}{ \begin{eqnarray}}
\newcommand{\ee}{\end{eqnarray}}
\begin{document}
\ \\
\vspace{3cm}

\begin{center}
{\Large Production of Dileptons in Heavy Ion Collisions at SPS-Energies\footnote{e-mail: vkoch@lbl.gov} 
}
\ \\
\ \\
\ \\
{\large Volker Koch}\\
Lawrence Berkeley National Laboratory, \\
Berkeley CA 94720, USA
\end{center}
\ \\
\ \\
\begin{center}
{\bf Abstract}
\end{center}
In this contribution we will discuss the production of low mass dileptons
in SPS-energy heavy ion collisions.
We briefly review the current theoretical situation before we turn to the
analysis of the recent data for Pb+Au. We also will discuss the role of baryons
as a source for dileptons. 
\ \\

\section{Introduction}
\begin{figure}[tbh]
\setlength{\epsfxsize}{0.7 \textwidth}
\centerline{\epsffile{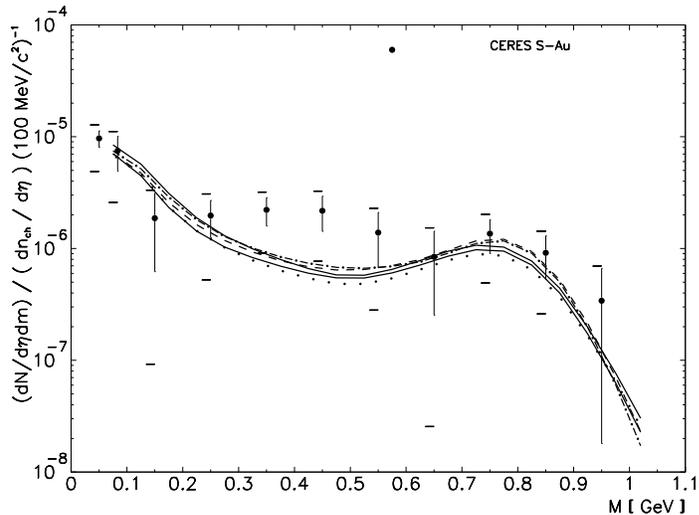}} 
\caption{Dilepton invariant mass spectrum for S+Au.  The curves are the
  results of \protect\cite{koch1} for different initial hadronic
  configurations. Data are from \protect\cite{CERES1}}
\label{fig:0}
\end{figure}

The study of low mass dileptons has recently received considerable interest. 
This has been triggered by the observation of enhanced production of dileptons
with
invariant mass around $400 - 500$~MeV in relativistic heavy ion collisions 
by the CERES collaboration \cite{CERES1,CERES2}. 
This enhancement has been studied in
various approaches, ranging from thermal model to complicated transport
models. All those calculations include the known hadronic decay channels into
lepton pairs and, in addition, dilepton production 
via re-interaction of
particles, most prominently pion annihilation. They find
that pion annihilation accounts for a large part of the observed enhancement,
while other channels such as the pion-rho scattering or the Dalitz decay of
the $\rm a_1$-meson are less important (see e.g. \cite{koch1,haglin}).
In ref. \cite{koch1} a large variety of initial conditions for the hadronic
fireball has been considered under the constraint that the final state
hadronic spectra are in agreement with experiment. Surprisingly little
variation has been found in the resulting dilepton spectra (see
fig. \ref{fig:0}). Certain initial
conditions would agree with the lower end of the sum of statistical and
systematic errors of the CERES data for the sulfur on gold reactions. In
\cite{koch1,koch2} in
medium modifications of pions and the pion nuclear from-factor in a pion gas
have been considered and have been found to be small. The conclusion of
\cite{koch1} and many other works (see \cite{ko_koch_rev} for a list of
references) is that in order to reach the central data points of the $\rm S +
Au$ measurement, additional in medium modifications need to be considered. 
Most of the attention received  the suggestion of Li et al. \cite{li1}, 
that a dropping of the mass of the $\rho$-meson with density, -- following,
with some modifications, the original conjecture of Brown and Rho \cite{br91} 
-- can
reproduce the central data points. On the other hand, Rapp et al. \cite{rapp} 
have extended the work of \cite{koch1,song} 
to include also the effect of baryons
for the in medium modification of pions. The present status of those
considerations is that the
in medium change of the pion dispersion relation leads only to a small
enhancement, whereas the inclusion of baryon resonances which couple  directly 
to
the rho meson appear to be able to increase the yield substantially. Most
important is here the $\rm N^*(1520)$ resonance, as first pointed out by the
Giessen group \cite{mosel}. The p-wave resonances, which have been first
considered by Friman and Pirner \cite{friman} appear to play a lesser role.
We should note, however, that the calculation of Steele et
al. \cite{steele,steele_new}, 
although similar in spirit, finds a much smaller effect due to baryons.

In this contribution we want to revisit the CERES data, in particular those for
the system $\rm Pb + Au$ \cite{CERES2}. These data have been analyzed to
provide not only an invariant mass spectrum but also transverse momentum
spectra and thus may give new insight into the relevant production mechanisms.
Recently, new (preliminary) results from the '96 run have been shown
\cite{lenkeit}. These data have much improved statistics as compared to the
published data from the '95 run.
We also will present arguments concerning the importance of baryons. According
the work of Rapp et al. baryons seem to be the most important source for the
low mass enhancement. In contradiction to that, our estimates in \cite{koch1}
found the baryons to be irrelevant.

\section{The $\rm Pb + Au$ data}
In this section we present some new results for the dilepton spectra for $\rm
Pb+Au$. The calculation is similar to that carried out in \cite{koch1} and we
refer to this reference for details. A new element is the inclusion of
the channel $\pi + \rho \rightarrow \pi + e^+ e^-$ \cite{pirho}. Using vector
dominance this process is related to the elastic $\pi + \rho \rightarrow \pi +
\rho$ scattering, which gives rise to the a collisional broadening of the rho
meson, as first discussed in \cite{haglin_coll}. We have
attempted to include the effect of the collisional broadening into our
transport model, by calculating the collisional width as a function of the
local pion density. This certainly is a crude method and needs to be refined in
the future. 
While there is some reduction of strength below the rho-omega peak due to the
collisional broadening of the rho, the overall effect is small and the
difference to the previous calculations \cite{koch1} are well within the
theoretical uncertainties as discussed in \cite{koch1}.

\begin{figure}[tbh]
\setlength{\epsfxsize}{0.6 \textwidth}
\centerline{\epsffile{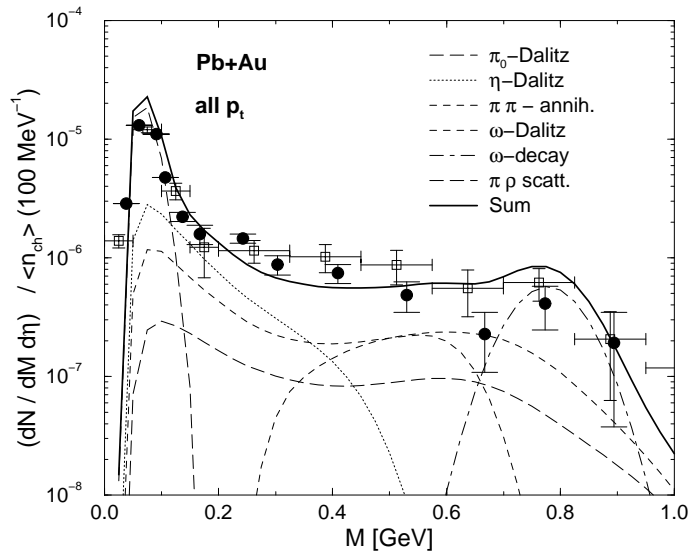}} 
\caption{Dilepton invariant mass spectrum  for semi central. 
The '95 data (open squares) are from  \protect\cite{CERES2} 
and the '96 data (full circles) are from 
\protect\cite{lenkeit}.
}
\label{fig:1}
\end{figure}

In fig. \ref{fig:1} we show the resulting invariant mass spectra 
together with the 
CERES data \cite{CERES2}, where only the statistical errors are shown. We have
also included the {\em preliminary} data from the '96  run (full circles)
\cite{lenkeit}. We find
a reasonable overall agreement, especially with the new, preliminary data.  

\begin{figure}[tbh]
\setlength{\epsfxsize}{0.48 \textwidth}
\centerline{\epsffile{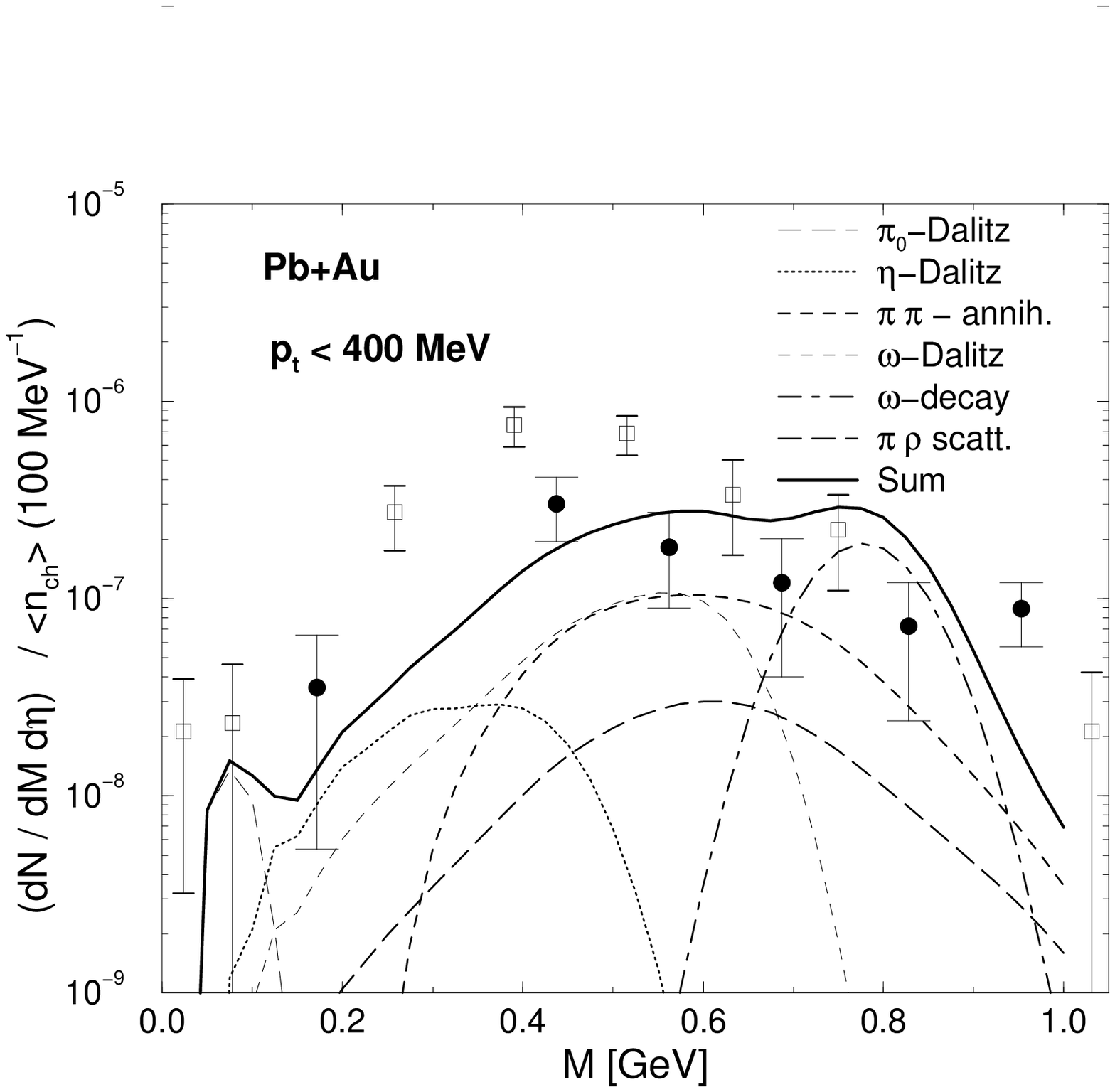} 
\setlength{\epsfxsize}{0.48 \textwidth}
\epsffile{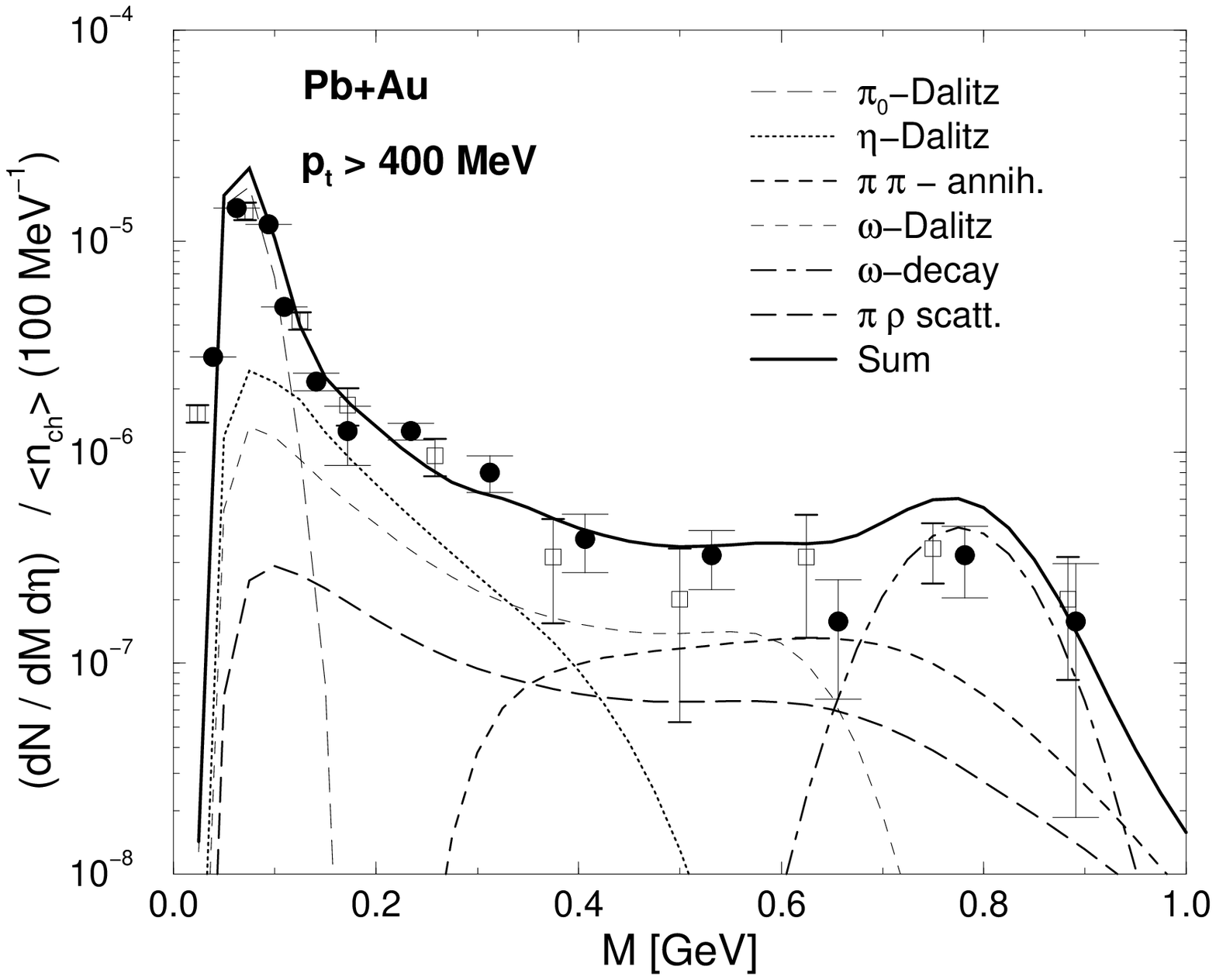} }
\caption{Dilepton invariant mass spectra for transverse momenta smaller
than 400~MeV (left panel) and larger than 400~MeV (right panel). 
The data are from  ref. \protect\cite{lenkeit}. Open squares are the '95 data
and full  circles are the '96 data. }
\label{fig:2}
\end{figure}

In fig. \ref{fig:2}  we show the invariant mass spectra 
for the tranverse momentum interval $p_t < 400 \rm \, MeV$ (left panel) and
$p_t > 400 \rm \, MeV$ (right panel). 
Again the agreement is quite reasonable. Some discrepancy might possibly be
around the rho-omega peak where our calculation overshoots the data somewhat.
However, as already pointed out in \cite{koch1} the strength around the
rho-omega peak is dominated by the omega decay contribution, which depends on
the abundance of omegas in the final state. This, however, is not very well
constrained by other data.

\begin{figure}[tbh]
\setlength{\epsfxsize}{0.5 \textwidth}
\centerline{\epsffile{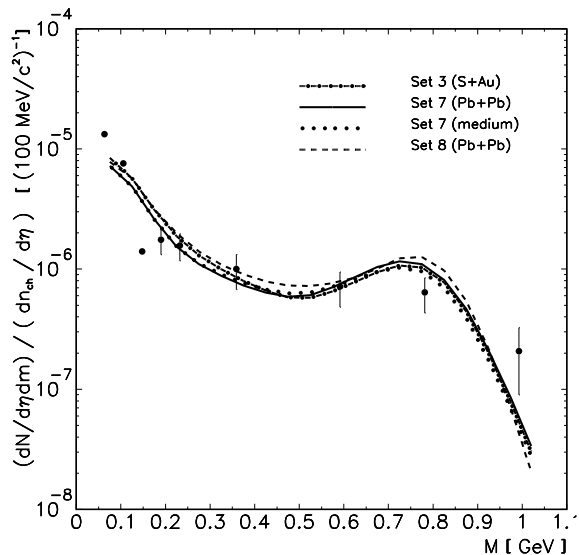} }
\caption{Dilepton invariant mass spectrum  central collisions. 
Curves are a prediction from \protect\cite{koch1}. 
The dotted line is the result including in medium correction in a pion gas. 
The data are 
from the '96 data set \protect\cite{lenkeit}.
}
\label{fig:2b}
\end{figure}

Finally in fig. \ref{fig:2b} we compare the {\em prediction} from \cite{koch1}
for central Pb+Au collisions with the central '96 data \cite{lenkeit}.
Again good agreement with the data is found. Also shown in this figure is the
invariant mass spectrum after in medium modification in a pion gas are taken
into account. These in medium modification also include the effects of chiral
symmetry restoration on the coupling of the photon to the rho meson
\cite{ioffe,song}. Notice, that it will be extremely difficult to extract
those effects from the data.  

To summarize this section, provided the still preliminary '96 data remain
unchanged it seems the the Pb+Au data are consistent with a simply hadronic
scenario without any large in medium modifications. As already pointed out,
in medium modification due to the presence of a pion gas are small and thus
their presence/absence can not be decided on from the present data. 
This situation could be improved considerably if the mass resolution would
allow to separate out the omega decay. This could then be `subtracted' and the
effects of chiral symmetry restoration should lower the remaining spectrum 
by about a factor of two around the rho-peak \cite{koch1}. The recently
completed upgrade of the CERES detector should allow for exactly that. 
But certainly the present Pb+Au data to not seem to call for 
sizable corrections as they were predicted by the dropping rho mass scenario
\cite{ko_qm}.

\section{The role of baryons}
\begin{figure}[tbh]
\setlength{\epsfxsize}{.8 \textwidth}
\centerline{\epsffile{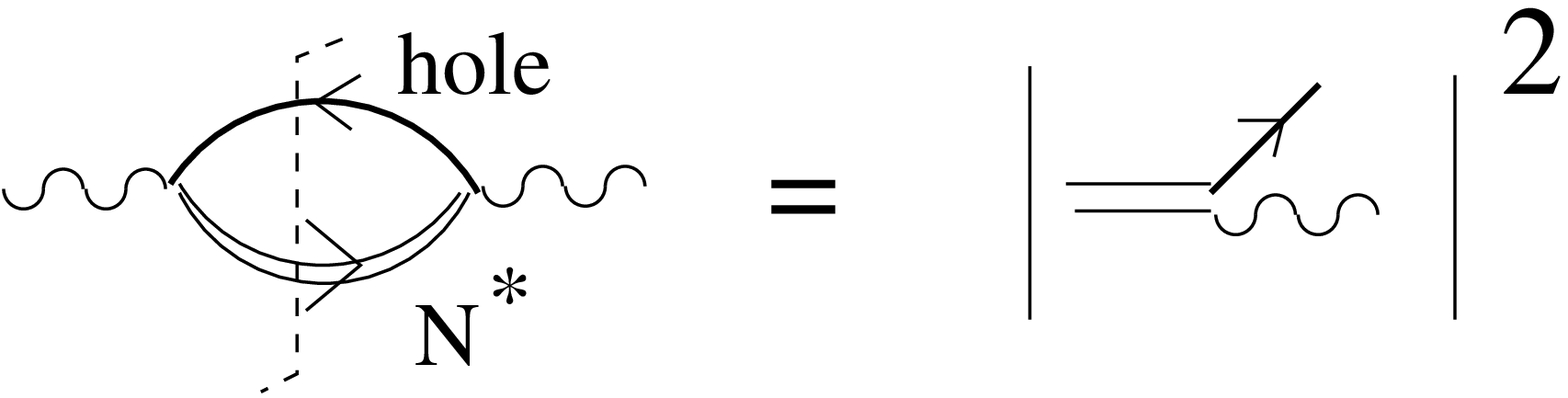}}
\label{fig:3}
\end{figure}

The work of Rapp et. al has emphasized the role of baryons as a possible
source for additional dileptons. They consider in medium modifications 
of the current-current correlator, the imaginary part of which is directly
related to the dilepton production rate. Following the suggestion of the
Giessen group \cite{mosel} Rapp et al. also find that the contribution of the
$N^*(1520)$-hole diagram, depicted in fig. \ref{fig:3} is the most important
one. However, as illustrated in fig. \ref{fig:3}, the imaginary part of this
diagram is nothing but the Dalitz - decay of the $N^*(1520)$. 

\begin{figure}[tbh]
\setlength{\epsfxsize}{.8 \textwidth}
\centerline{\epsffile{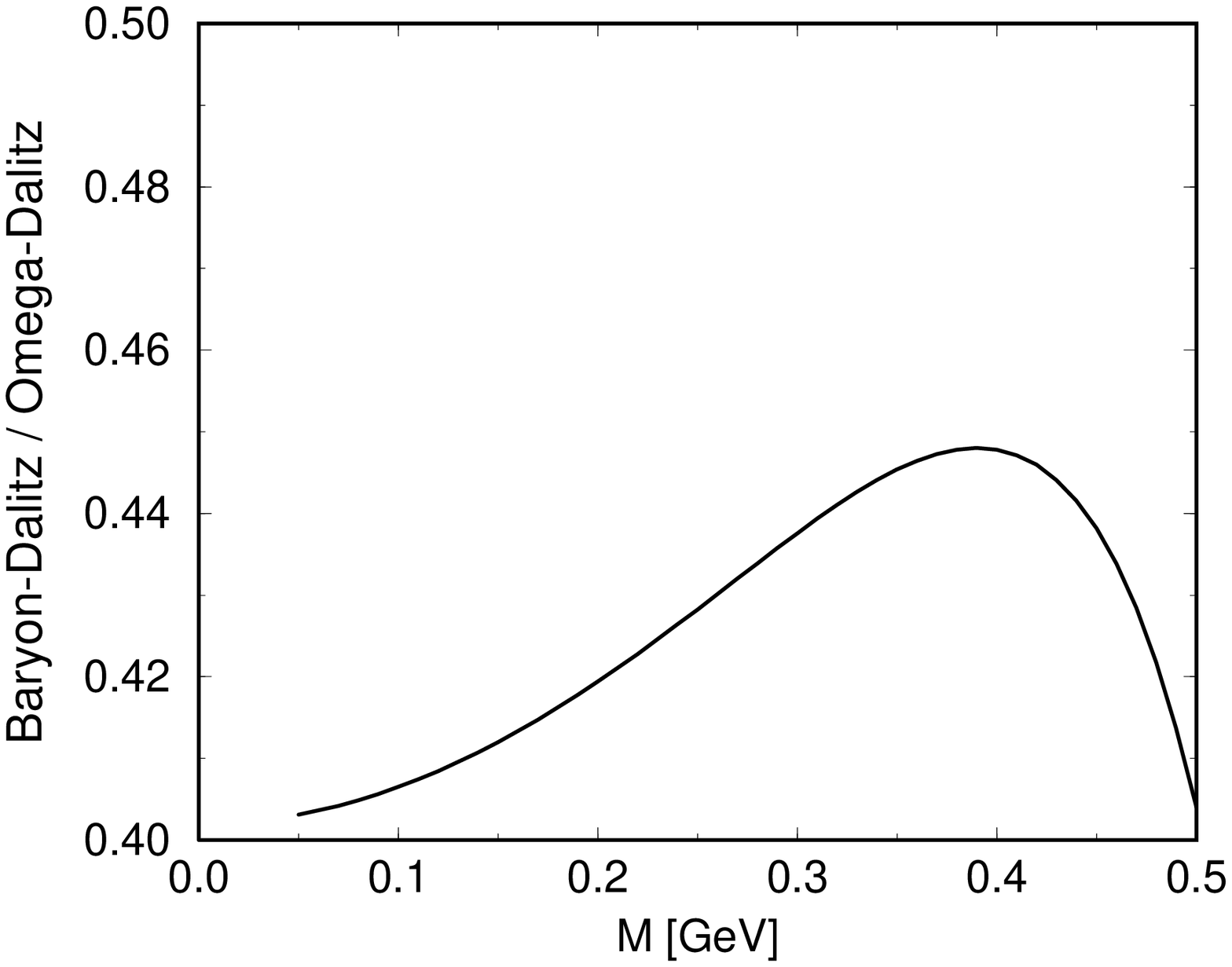}}
\caption{Ratio of baryon-Dalitz decay over $\omega$-Dalitz decay}
\label{fig:4}
\end{figure}

The contribution of the Dalitz decays of baryons has already been estimated
in ref. \cite{koch1}. In this estimate, the formula for the branching ratio of
photon to Dalitz decay of
the Delta \cite{wolf} has been extended to higher masses. In order to arrive
at an conservative estimate of an upper limit the fraction of higher lying
resonances has purposely been overestimated by a factor of two. The
photon decay width has been chosen to be $1 \, \rm MeV$, which again is on the
large side. The resulting dilepton spectrum has then been compared with that
of the omega, in order to minimize the  effect of the detector acceptance. 
The ratio of these yields is shown in fig. \ref{fig:4}. The baryons hardly
contribute half as much as the omega Dalitz. Considering the contribution of
the omega Dalitz as shown in fig. \ref{fig:1} it seems that the baryons are
anything but irrelevant.

So what is the difference between this estimate and the results of Rapp et al.
Several possibilities come to mind:
\begin{itemize}
\item The $N^*(1520)$ has a considerably larger Dalitz decay width than the
the extrapolation from the Delta decay width would predict. This might be
possible, in particular, because the $N^*(1520)$ couples rather strongly to
the $\rho$-meson. This is presently being investigated \cite{gale} and we find
that in both a relativistic as well as a nonrelativistic description the Dalitz
decay is well in line with fig. \ref{fig:4}. Similar results are also found by
\cite{steele_new}.
\item Rapp et al. sum the RPA-type Dyson-series for these diagrams. So in
  principle there could be collective effects, which are ignored in the simple
  calculation of the Dalitz decay. It appears, however,  rather 
  unlikely that at the
  temperatures under consideration, collectivity can play an important role.
\item
  Another source of discrepancy is the  baryon to pion ratio assumed in the
  calculations. Using the freeze out parameters of \cite{rapp} we find a pion 
  to baryon ratio for $Pb+Au$ of about 3,   
  whereas a ratio of close to 6 is observed in experiments. 
  The estimate of \cite{koch1}, on the other hand was based on a realistic
  pion/baryon ratio. 
\end{itemize}

All this points are specific to the environment created in a CERN energy heavy
ion collision. At lower energies or in proton/pion nuclear reactions the
density effects could very well be large. This will be investigated in the
near future by the HADES detector at the GSI.

\section{Conclusions}
At this point it is very hard to draw any firm conclusions as the new CERES
data are still preliminary.  Taking these data at face
value, however,  it appears that
no or only small in medium corrections are needed in order to explain the 
data. This would be
somehow unfortunate, although, as shown in \cite{koch1}  in medium
modifications due to the presence of pions indeed  give rise  only to small 
corrections. Certainly, all these calculations are
rather unconstraint. For instance the number of omegas can be chosen within a
considerably wide range. While these uncertainties have already been addressed,
a measurement with a mass resolution  which is sufficient to constrain the
number of omegas in the final state would reduce these uncertainties to a
large extent. This would then  provide the basis for the search for the 
more subtle effects which, one would think, should be there.

\section{Acknowledgments}
I would like to thank my collaborators C.M. Ko,  
C. Gale and  A. Kanti Dutt-Mazumder with whom I am
presently working the the issues addressed in this contribution. 
This work was supported by the Director, 
Office of Energy Research, Office of High Energy and Nuclear Physics, 
Division of Nuclear Physics, and by the Office of Basic Energy
Sciences, Division of Nuclear Sciences, of the U.S. Department of Energy 
under Contract No. DE-AC03-76SF00098.

\end{document}